# Methylamine Vapor Exposure for Improved Morphology and Stability of Cesium-Methylammonium Lead Halide Perovskite Thin-Films

Akash Singh[1,2,*], Arun Singh Chouhan[1] and Sushobhan Avasthi[1]

[1]Centre for Nano Science and Engineering, Indian Institute of Science, Bangalore, India, 560012

[2]PDPM Indian Institute of Information Technology, Design and Manufacturing, Jabalpur, India, 482005

akashsingh@iisc.ac.in

**Abstract.** Mixed-cation Cesium-Methylammonium lead halide perovskite ($Cs_xMA_{1-x}PbI_{3-x}Br_x$) thin-films have been used to demonstrate stable and efficient perovskite devices. However, a systematic study of the Cs incorporation on the properties of the perovskite films has not been reported. In this report, Impact of Cesium incorporation on the minority carrier recombination lifetime of Cesium-Methylammonium lead halide perovskite thin-films is studied. The lifetime for the as-deposited perovskite films decreases with increasing concentration of cesium. However, mixed cation perovskite film is more stable, showing higher lifetime (15-20 µs) after 9 hours of ambient exposure than just after deposition (6-13 µs). 'Methylamine Vapor Exposure' (MVE) technique was used to improve the morphology of the as-deposited film. MVE treated films are more oriented along (110) direction and were even more stable in ambient, with $Cs_{0.10}MA_{0.90}PbI_{2.90}Br_{0.10}$ films showing lifetime of almost 50 µs after 9 hours of ambient exposure, twice the lifetime of a comparable $MAPbI_3$ film. These results throw light on why mixed-cation cesium-methylamine lead halide perovskite films are better for highly efficient and stable perovskite solar cells.

**Keywords:** Carrier Lifetime, Methylamine Vapor Exposure, Perovskites thin films, Photovoltaics, Stability

## 1 INTRODUCTION

Organic-inorganic hybrid perovskites are very interesting materials for photovoltaic and optoelectronic applications, owing to their excellent opto-electrical properties like high absorption coefficient, tunable direct band gap, lower excitonic binding energies, defect tolerance and long carrier recombination lifetimes. State-of-the-art hybrid perovskite solar cells yield efficiencies as high as 22% [1]. Methylammonium lead iodide ($MAPbI_3$) is the most well-studied perovskite absorber layer, however it tends to degrade in ambient conditions due to the hygroscopic nature of methylammonium cations [2]. One way to limit degradation is to completely replace the Methylammonium cation with inorganic cesium, forming inorganic cesium lead iodide ($CsPbI_3$) perovskite. Unfortunately, at room temperature $CsPbI_3$ exists in orthorhombic phase which is yellow in color and has poor photovoltaic properties. Black-colored cubic phase of $CsPbI_3$, which shows the desirable photovoltaic properties, exists only at temperatures higher than 300°C [3]. To solve this problem, researchers have investigated mixed cation perovskite formulation, where 10-15% of methylamine is replaced with Cs in the perovskite film to form $Cs_xMA_{1-x}PbI_{3-x}Br_x$. This has already been shown to yield device with good efficiency and improved the stability, but a systematic study on the effect of cesium incorporation on the material properties, such as carrier recombination lifetimes has not been done. In this work, the effect of Cs incorporation on stability and performance of mixed cation perovskite films ($Cs_xMA_{1-x}PbI_{3-x}Br_x$) is studied by measuring the minority carrier recombination lifetime, at varying level of Cs incorporation (x in %). Furthermore, Methylamine Vapor Exposure (MVE) has been shown to improve the morphology of $MAPbI_3$ films, leading towards increase in efficiency and stability of devices [4]. However, the suitability of MVE technique on mixed cation perovskite films and its effect on their stability has not been studied before in detail. This work also shows the suitability of this technique in improving the morphology, crystallinity, lifetime values and stability of the $Cs_xMA_{1-x}PbI_{3-x}Br_x$ films.

## 2    EXPERIMENTAL DETAILS

Mixed cation perovskite precursor solutions (1M) containing 0, 5, 10 and 15% Cesium were prepared using different stoichiometric ratios of Cesium Bromide (CsBr), methylammonium iodide (MAI) and Lead Iodide ($PbI_2$) in N-N dimethylformamide (DMF) solvent. The prepared solution was stirred overnight at 100°C. Glass wafers (RMS roughness ~ 5nm) post dicing were cleaned by successive ultra-sonication in acetone, iso-propanol and deionized water for 20 minutes each, followed by drying under a nitrogen handgun. Cleaned glass wafers were placed inside a nitrogen glove-box with oxygen and water content below 0.1 ppm. Glass-wafers were pre-heated for 30 minutes at 100 °C and perovskite solution was spin-coated at 2000 rpm for 30 seconds. Spin-coated films were further annealed for 20 minutes at 100°C. For MVE treatment, thin-films were exposed to methylamine vapour for ~1 second at room temperature and allowed to recrystallize at room-temperature [4].

Minority carrier recombination lifetime values were measured for perovskite thin films by Freiberg Instrument MDP-spotusing a 405 nm laser(180 mW) [4,5]. Perovskite thin films were characterized using Rigaku Smart lab X-ray Diffractometer Cu-K-alpha radiation. Morphological study was carried out using Scanning Electron Microscopy ZEISS Ultra 55.

# 3   RESULTS AND DISCUSSION

Minority carrier recombination lifetime is a measure of the duration for which photo generated carriers exist in a material, before being lost to recombination. Recombination lifetime directly depends on the energy-position and density of defects and trap states present in the material. In our study, microwave detected photoconductivity (MDP) was used to characterize the carrier recombination lifetime. It is a non-invasive technique wherein a pulsed laser source excites the charge carriers, which temporarily increases the conductivity of the perovskite film. As the excess photo generated carriers recombine, the conductivity reverts to the equilibrium value. The conductivity can be measured in a non-contact mode using microwave because microwave reflectivity off a semiconductor surface is a function of the conductivity of a film. By continuously probing the sample with microwaves, the decay rate of photo generated carriers can be extracted. The lifetime so measured is an "effective" number that includes both the bulk and surface recombination processes that can occur in a film. Assuming, the surface recombination does not change with minor changes in material composition, the effective lifetime can be assumed to be proxy for the bulk recombination lifetimes, and hence a measure of material quality. In this work we used the effective carrier recombination lifetime measurements (henceforth called carrier lifetime) on a series of $Cs_xMA_{1-x}PbI_{3-x}Br_x$ films with different levels of Cs-incorporation (x = 0, 0.05, 0.10 and 0.15). Cesium incorporation was not increased beyond x=0.15, because higher concentration of Cs lead to segregation, where Cesium and methylamine perovskites crystallize separately [6]. Carrier concentration decay profile, measured using MDP-Spot, of the freshly prepared mixed cation perovskite films with 0, 5, 10 and 15 % of Cesium content is shown in Fig.1 (a). In general higher the lifetime, slower the decay. The decay can be fit to an exponential, the characteristic depth of which is the carrier lifetime. Extracted carrier lifetime of the as-deposited samples was 21.5 µs for pure MA perovskite film (0% Cs). As the cesium concentration increased from 0 to 15%, the lifetime fell from 21.5 to 5.8 µs. This gradual decrement in the lifetime values could be attributed to the increase in the inhomogeneity due to Cesium incorporation along with the retarded grain growth where Cesium messes up with the crystallization of perovskite.

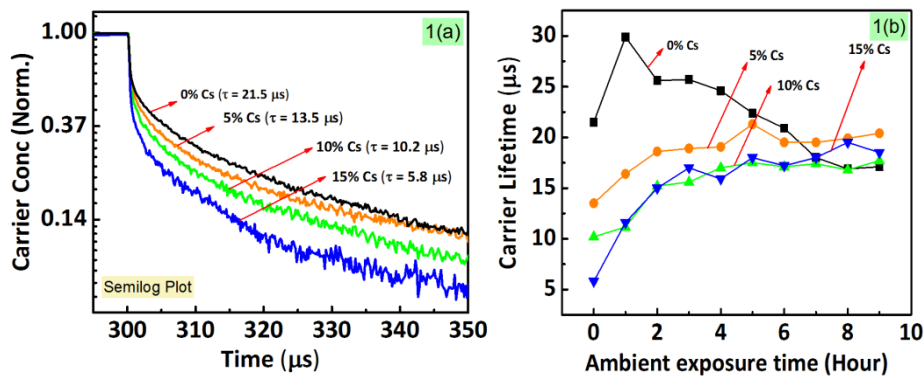

**Fig1.(a)** Carrier density decay in as-deposited perovskite films with different amount of Cs **(b)** Evolution of carrier lifetime values with exposure to ambient

Carrier lifetime closely correlates with the film quality and ultimately to the efficiency of the solar cell. So, we used carrier lifetime to characterize the stability of the films. The carrier lifetimes of the perovskite films with different amount of Cesium were measured regularly over a period of 9 hours, during which the films were stored in air, exposed to ambient humidity and oxygen. Fig. 1 (b) shows the evolution of carrier lifetime values over a period of 9 hours of ambient ageing. In the first one hour, films with 0% Cs showed an increase in the carrier lifetime value from 21 µs to 30 µs [4]. This has been observed before, and typically is attributed to the healing behavior of the moisture present in the ambient environment, which facilitates better carrier transport across grains and reduced recombination [7]. After one hour of exposure, the lifetime values gradually decrease over time, showing the degradation in the films. Cesium incorporated films (x>5%) also show an increase in lifetime with ambient exposure, probably due to the similar healing behavior in the ambient. However, the notable difference is that the lifetime continues to increase even 9 hours after the exposure, with all the composition showing lifetime of 16-20 µs. This suggests that compared to pure MA perovskite films, the Cs-incorporated films are more resilient to degradation. The observation supports previous reports which demonstrated increase stability in mixed-cation perovskite solar cells.

The carrier lifetime has a direct correlation with the morphology in terms of grain sizes and the boundaries associated with it. Grain boundaries act as non-radiative recombination centers leading towards faster decay of photo generated carriers. Smaller grains will lead to more carrier recombination, lower carrier lifetime, and lower device performance. In order to probe the effect of Cs-incorporation on the morphology of the deposited films, we carried out the morphological study of the as-deposited 10% Cs containing perovskite film. Fig 2 (a) shows the SEM image of the as-deposited perovskite film. The films were non-uniform with poor surface coverage, having micron sized pin-holes and grain sizes in the range of 300-500 nm. Films with such morphology will lead to low efficiency solar cells. Solvent-solvent extraction and post treatments are widely used to improve the morphology of the perovskite films. It has been previously shown that exposure to methylamine vapors (MVE) improves the MAPbI3 film quality by means of complex formation and perovskite recrystallization [4]. However, there are no reports of MVE treatment applied to mixed-cation perovskite films. To investigate, we exposed mixed cation perovskite film with 10% Cs to MA vapor for ~1 second. The exposure led to the formation of an optically bleached frustrated complex. Once the vapors were removed from the ambient, the complex immediately decomposed, forming shiny black film of mixed-cation perovskite. The resulting "MVE" film showed excellent coverage with no observable pin holes. The grain-size, or more accurately the domain size, also increased to 1-3 microns (Fig 2(b)).

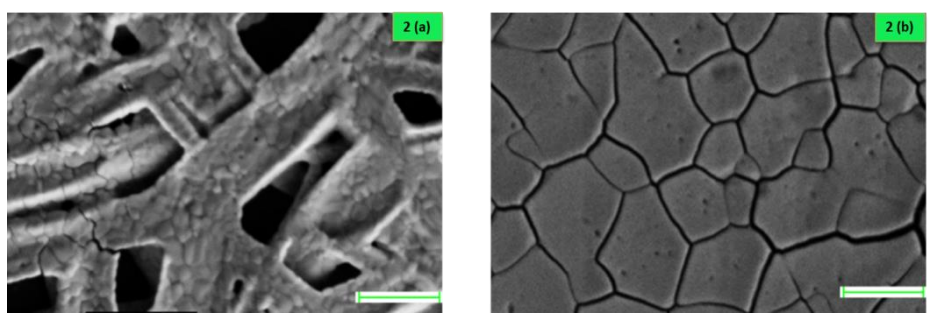

**Fig. 2.** SEM image of **(a)** as-deposited 10% Cs containing perovskite film **(b)** MVE treated 10% Cs containing perovskite film. Scale bar is 2 micron

The test proved that MVE technique can be used to improve the film quality even for mixed cation perovskite films, a novel result. To check the phase purity we recorded the XRD pattern of the 10% Cs containing as-deposited and MVE treated perovskite film. Fig. 3 shows the normalized grazing angle XRD pattern of both the films. Both the as-deposited and MVE treated films are phase pure perovskites, with no impurity peaks of CsPbI3/CsPbBr3 perovskite. The MVE treated films were more oriented in the (110) crystallographic direction with reduction in the intensities of other planes. Further, the full width half maximum (FWHM) value corresponding to the perovskite primary peak at ~ 14° reduced to 0.376° in MVE treated films as compared to 0.402° in as-deposited films confirming the enhancement in the crystallite grain size.

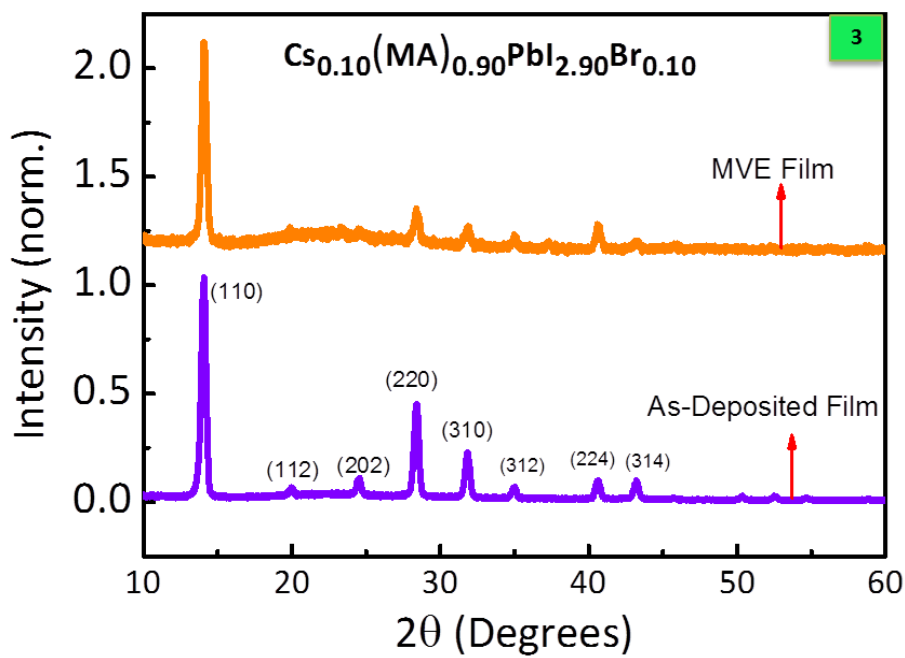

**Fig. 3.** XRD patterns of 10% Cs containing as-deposited and MVE treated perovskite films

To investigate the change in carrier lifetime values owing to the MVE treatment, we further carried out the carrier lifetime measurements. The obtained value showed a significant increase in lifetime values for both the 0% and 10% Cs containing MVE treated films. Fig 4 (a). shows the carrier concentration decay curve for 0% and 10% Cs containing as-deposited (dotted lines) and MVE treated (continuous lines) films. The carrier lifetime values for the freshly prepared 0% Cs films increased to 31 µs from 21.5 µs after MVE treatment and similar results were observed for 10% Cs incorporated films whereby lifetime values increased to 25 µs from 10 µs. This significant increase in lifetime values was attributed to increased grain size with reduced number of grain boundaries after MVE treatment (As shown in Fig 2). Although MVE treatment lead to an increase in the lifetime values for both the 0% and 10% Cs incorporated film, but the obtained values were higher in non-cesium containing films. This decrement was well in agreement with the fact that Cesium incorporation retards the crystal growth leading towards smaller grains. We compared the grain sizes of the MVE treated 0% Cs and 10% Cs containing perovskite films and found the same results were non-Cesium containing film had grain sizes in the range of 4-6 microns whereas films with 10% Cs had grain sizes of 1-3 microns.

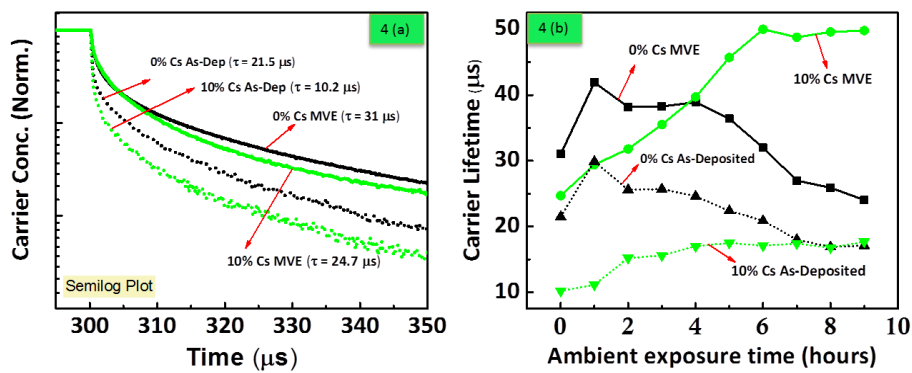

**Fig. 4.** Carrier concentration decay curve for as-deposited and MVE treated perovskite films with 0 and 10% Cs **(b)** Evolution of carrier lifetime values with exposure to ambient

To probe further, we once again used the carrier lifetime measurement as a measure to study the stability of MVE treated films. Fig 4 (b) shows the carrier lifetime evolution profile for the as-deposited and MVE treated perovskite films with 0 and 10% of Cs incorporation for 9 hours of continued ambient exposure. As can be inferred from the figure, the carrier lifetime for the MVE treated films were always significantly higher than the as-deposited films. 0% Cs films after MVE treatment did show an increment in the lifetime value which increased from 31 μs to 42 μs over an hour but soon after that, degradation dominated over moisture assisted healing behavior resulting into gradual decrement in the lifetime values on further ambient exposure. Whereas, for the films with 10% Cs content, the lifetime values increased gradually from 25 μs to 50 μs in 5-6 hours and saturated over there showing no further decrement. This suggested that the 10% Cs containing MVE treated films were not only stable but also show improvement in its photo-physical properties rendering higher recombination lifetimes which would lead towards better performing devices with improved carrier collection.

## 4    Conclusion

We reported that Cs incorporation results in a monotonic decrease in carrier recombination lifetime of $Cs_xMA_{1-x}PbI_{3-x}Br_x$ films. However, Cs-incorporated films are more stable, with recombination lifetime increasing by 40-300% for different Cesium concentration over 9 hours of ambient exposure. MVE technique also works on mixed cation perovskite films which improve the morphology and quality of the as-deposited films, leading to higher recombination lifetimes and resilience to ambient exposure degradation. This work shows the importance of the moisture assisted healing behavior on mixed cation perovskite films. Devices, if fabricated in controlled humid environment using mixed cation perovskite film as absorber layer will eventually leads to higher carrier collection efficiencies.

## Acknowledgements

This work was supported under the U.S.−India Partnership to Advance Clean Energy-Research (PACE-R) for the Solar Energy Research Institute for India and the United States (SERIIUS), funded jointly by the U.S. Department of Energy (Office of Science, Office of Basic Energy Sciences, and Energy Efficiency and Renewable Energy, Solar Energy Technology Program, under Subcontract DE-AC36-08GO28308 to the National Renewable Energy Laboratory, Golden, Colorado) and the Government of India, through the Department of Science and Technology under Subcontract IUSSTF/JCERDC-SERIIUS/2012 dated 22nd Nov. 2012. The work is also supported by Department of Science and Technology (DST), Government of India, under project reference no: SB/S3/EECE/0163/2014. The last two authors would also like to acknowledge the support from PhD Fellowship and Young Faculty Research Fellowship under the Visvesvaraya PhD Scheme for Electronics and IT by Ministry of Electronics & Information Technology (MeitY), Government of India.